\documentclass[twocolumn,showpacs,showkeys,prl,aps]{revtex4}
\usepackage{graphicx}

\newcommand{ \be }{\begin{equation}}
\newcommand{ \ee }{\end{equation}}
\newcommand{ \bea }{\begin{eqnarray}}
\newcommand{ \eea }{\end{eqnarray}}

\begin{document}

\title{Long Range Correlations and the Soft Ridge 
in Relativistic Nuclear Collisions}

\author{Sean Gavin,$^a$ Larry McLerran,$^b$ and George Moschelli$^a$ }
\affiliation{
a) Department of Physics and Astronomy, Wayne State University, 666
W Hancock, Detroit, MI, 48202, USA\\
b)  Physics Department and RIKEN Brookhaven Research Center, Building 510A 
Brookhaven National Laboratory, Upton, NY, 11973, USA}

\date{\today}
\begin{abstract}
Relativistic Heavy Ion Collider experiments  exhibit correlations peaked in relative azimuthal angle and extended in rapidity.  Called the ridge, this peak occurs both with and without a jet trigger.  We argue that the untriggered ridge arises when particles formed by flux tubes in an early Glasma stage later manifest transverse flow.  Combining a blast wave model of flow fixed by single-particle spectra with a simple description of the Glasma, we find excellent agreement with current data. 
\end{abstract}


\maketitle

Correlation measurements of high transverse momentum particles provided the first striking experimental evidence of jet quenching.  Further studies of the correlation of high $p_t$ particles with more typical particles reveal a complex pattern of correlations as a function of relative pseudorapidity $\eta = \eta_1-\eta_2$ and azimuthal angle $\phi = \phi_1-\phi_2$. In particular, a ``hard ridge'' of enhanced correlation is observed near $\eta = \phi = 0$  that is narrow in $\phi$ and broad in $\eta$ \cite{Putschke:2007mi}.
Interestingly, the STAR collaboration reports a similar ridge in correlations of particles of any $p_t$, i.e., without a jet trigger \cite{starDense,ron,Daugherity:2006hz}.  As with the hard ridge, the width of the untriggered soft ridge is broad in $\eta$ and narrow in $\phi$.  Preliminary PHOBOS data suggests that the hard ridge and possibly the soft ridge may extend over the broad range $-4 < \eta < 2$ \cite{Wenger:2008ts}. 

We argue that the soft ridge is a consequence of early-stage rapidity correlations in concert with late-stage transverse flow. Correlations over several rapidity units can only originate at the earliest stages of an ion collision when the  first partons are produced \cite{larry}. Hydrodynamics and other later-stage effects can modify these correlations, but are limited to a horizon of $\sim 1-2$ rapidity units. Analogous to super-horizon fluctuations in the cosmos, 
these long range correlations can therefore reveal the `little bang'  in each nuclear collision at its birth. 

Almost instantaneously after a collision of two nuclei, Color Glass Condensate (CGC) theory predicts that the transverse fields of each nucleus are transformed into longitudinal fields that are approximately uniform in rapidity \cite{mv,lappi}. The fields, which are random over transverse distances larger than $Q_s^{-1}$ where $Q_s\sim 1-2$~GeV is the saturation scale, comprise a series of flux tubes. Long range rapidity correlations arise because particles from the same flux tube start at nearly the same transverse position, regardless of rapidity.
 
Pressure builds as the flux tubes fragment to form particles. The resulting transverse flow modifies these long range correlations by focusing particles into a narrow region in $\phi$. Suppose that the transverse fluid velocity has the Hubble-like form $\gamma_t\mathbf{v}_t = \lambda \mathbf{r}_t$.
A fluid cell a distance $r_t$ from the center of the collision volume will then have a mean speed $v_t$. When this cell freezes out, it releases particles into an opening angle $\phi\sim v_{th}/v_t \propto (\lambda r_t)^{-1}$, for a thermal velocity  $v_{th}\sim 1$.  This modifies the correlations, since particles near the center of the collision volume have a large opening angle, while those from a larger $r_t$ have a smaller $\phi$.
Voloshin has long stressed the connection between flow and $p_t$ correlations  \cite{Voloshin:2003ud}.

In this letter we focus on the centrality dependence of the height and azimuthal width of the near side peak of the soft ridge. We will see that this dependence can be explained using CGC-Glasma scaling arguments combined with blast wave calculations. However, we remark that the measured correlation function is not a plateau in rapidity, but a broad structure of width $\sim 1-2$ units, perhaps with tails extending higher in rapidity \cite{Daugherity:2006hz,Wenger:2008ts}. We will address this rapidity dependence elsewhere, since that analysis requires quantum corrections to the Glasma \cite{larry} as well as viscous corrections to the hydrodynamic treatment \cite{gavin06}. We will also not discuss the jet triggered data, which would require a description of the passage of the jet through the high-density environment produced by the nuclear collision \cite{hardRidge,Shuryak:2007fu}. 

Flux tubes arise naturally in descriptions of high energy collisions \cite{lownusinov}. The Glasma description incorporates many of these features, and in a high density environment such as is produced by the collisions of nuclei, allows  for a systematic weak coupling computation. The contribution of flux tubes to long range correlations is studied in the Glasma formulation in ref.~\cite{larry}. 
We imagine the Glasma to be filled with flux tubes of large longitudinal extent but small transverse size $\sim Q_s^{-1}$. Each flux tube yields a multiplicity of  $\sim \alpha_s(Q_s)^{-1}$ gluons. The number of flux tubes is proportional to the transverse area $R_A^2$ divided by the area per flux tube, $Q_s^{-2}$.  The rapidity density of gluons is therefore 
\begin{equation}
{{dN}/{dy}} \sim {\alpha_s}^{-1}Q_s^2R_A^2.
\label{eq:Nscale}
\end{equation} 
The number of final hadrons scales similarly \cite{Kharzeev:2000ph}.

We characterize correlations in the Glasma and later in the evolution using the spatial correlation function 
\begin{equation}
c(\mathbf{x}_1, \mathbf{x}_2) = n_2(\mathbf{x}_1,\mathbf{x}_2) - n_1(\mathbf{x}_1) n_1(\mathbf{x}_2),
\label{eq:CorrDef}
\end{equation}
where $n_1$ and $n_2$ are the single and pair densities.  In the absence of correlations, $ n_2(\mathbf{x}_1,\mathbf{x}_2) \rightarrow n_1(\mathbf{x}_1) n_1(\mathbf{x}_2)$ so that $c$
vanishes. The integral $n_2$ over both positions gives the number of pairs averaged over events 
$\langle N(N-1)\rangle$. When correlations are negligible, the integral of $c$ 
vanishes -- as it must -- because $N$ follows Poisson statistics and, therefore,  $\langle N(N-1)\rangle\rightarrow \langle N\rangle^2$. 

We take pairs from the same flux tube as correlated and neglect correlations between tubes. Furthermore, we assume that correlations are independent of rapidity. The correlation function then depends only on the relative transverse position  $\mathbf{r}_t = \mathbf{r}_{1,\, t} -  \mathbf{r}_{1,\, t}$ as well as the average  $\mathbf{R}_t = (\mathbf{r}_{1,\, t} +  \mathbf{r}_{1,\, t})/2$. The correlation length in $r_t$ is roughly the flux tube size $\sim Q_s^{-1}$, while the correlation length in $R_t$ of order of the transverse system size $R_A$. For $Q_s^{-1} \ll R_A$ we take the correlation function to be point-like in $r_t$ and broad in $R_t$, writing 
\begin{equation}
c(\mathbf{x}_1, \mathbf{x}_2)
 = {\cal R}\,\delta(\mathbf{r}_t ) \rho_{{}_{FT}} (\mathbf{R}_t).
\label{eq:param}
\end{equation}
Here, $\rho_{{}_{FT}}(\mathbf{R}_t)$ describes the transverse distribution of flux tubes in the collision volume, which we assume follows the thickness function of the colliding nuclei
\be
          \rho_{{}_{FT}}(\mathbf{R}_t) =  {{ 2 \langle N\rangle^2} \over {\pi R^2_A}}  \left(1 - {{R_t^2}\over{R_A^2}}\right)
\ee          
for $R_t \leq R_A$, and zero otherwise. 
Integrating both sides of eq.~(\ref{eq:param}) with respect to $r_t$ and $R_t$, we find
\begin{equation}
\langle N\rangle^2{\cal R} = \int c\, d^3x_1 d^3x_2 = \langle N^2\rangle - \langle N\rangle^2 - \langle N\rangle.
\label{eq:varianceDef}
\end{equation}
To see how $\cal R$ depends on $Q_s$, think of each flux tube as a source that produces particles with a mean multiplicity $\mu$ and variance $\sigma^2$. For $K$ flux tubes, the mean multiplicity is $\mu K$ and the variance is $\sigma^2 K$. If $K$ fluctuates from event to event, then the mean multiplicity is $\mu \langle K\rangle$ and the variance is $\sigma^2 \langle K\rangle + \mu^2 (\langle K^2\rangle - \langle K\rangle^2)$. Therefore 
\begin{equation}
{\cal R} ={{\sigma^2-\mu}\over {\mu^2}}{{1}\over{\langle K\rangle}} + {{\langle K^2\rangle - \langle K\rangle^2}\over{\langle K\rangle^2}}.
\label{eq:source}
\end{equation}
Particle production from a flux tube is a Poisson process, since the flux tube is a coherent state. It follows that $\sigma^2=\mu$, so that the first contribution vanishes. For large $K$, the second term is $\propto \langle K\rangle^{-1}$.  

We combine these results to obtain a scaling relation for the integrated strength of correlations in the Glasma 
\begin{equation}
{\cal R}\propto \langle K\rangle^{-1} = (Q_sR)^{-2},
\label{eq:Rscale}\end{equation}
a result supported by momentum-space calculations in \cite{larry}. In contrast, the mean multiplicity in a rapidity interval scales as $\alpha_s(Q_s)^{-1}Q_s^2R^2$; see eq.~(\ref{eq:Nscale}). This difference will prove significant later. We comment that  (\ref{eq:Rscale}) and similar CGC relations may not quantitatively describe $pp$ or peripheral collisions at the energies studied here, although phenomenological string models may apply.

We now turn to discuss the impact of these long range correlations on the final-state particle correlations. As the partons emitted from these flux tubes locally equilibrate, transverse flow builds. To describe the effect of thermalization and flow on the pair correlation function at freeze out, 
we generalize the common blast-wave model 
\cite{schnedermann,
Kiyomichi:2005va, Barannikova:2004rp, Iordanova:2007vw, Retiere:2003kf}. To begin, recall that the Cooper-Frye single-particle distribution is 
\begin{equation}
 \rho_1(\mathbf{p}) \equiv dN/dyd^2p_t = 
\int f(\mathbf{x},\mathbf{p}) \, d\Gamma,
\label{eq:singles}
\end{equation}
where $f(\mathbf{x},\mathbf{p}) =  {(2\pi)^{-3}}\exp\{-p^\mu u_\mu/T\}$ is the Boltzmann phase-space density for a temperature $T$ and fluid four-velocity $u_\mu$, and $d\Gamma = p^\mu d\sigma_{\mu}$ is the element of flux through the four dimensional freeze out surface along which particle interactions effectively cease.  We assume that freeze out occurs at a proper time $\tau_F$, so that $p^\mu d\sigma_\mu = \tau_F m_t \cosh(y-\eta)d\eta d^2 r_t$, where $\eta = (1/2)\ln((t+z)/(t-z))$ is the spatial rapidity. We follow ref.~\cite{schnedermann} and write the four velocity of the longitudinal-boost invariant blast wave as $u_\mu = \gamma_t (\cosh\eta, \mathbf{v}_t, \sinh\eta)$, where $v_t$ is the transverse velocity and $\gamma_t = (1-v_t^2)^{-1/2}$. The phase space density is then $f \propto \exp\{-\gamma_t m_t \cosh(y-\eta)/T\}  \exp\{\gamma_t \mathbf{v}_t \cdot \mathbf{p}_t/T\}$. We take the transverse velocity to be
$\gamma_t \mathbf{v}_t \approx \lambda \mathbf{r}_t$,
a widely-used ansatz that adequately describes much of SPS and RHIC data. The calculation of (\ref{eq:singles}) is standard and follows \cite{schnedermann}.

To exhibit the effect of flow on particle correlations, we use the momentum-space correlation function 
\begin{equation}
r (\mathbf{p}_1, \mathbf{p}_2) = \rho_2 (\mathbf{p}_1, \mathbf{p}_2) - \rho_1(\mathbf{p}_1)\rho_1( \mathbf{p}_2) 
\label{eq:MomCorr0}
\end{equation}
where $\rho_2 (\mathbf{p}_1, \mathbf{p}_2) = dN/dy_1d^2p_{t1}dy_2d^2p_{t2}$ is the pair distribution.  Generalizing (\ref{eq:singles}), we write
\begin{equation}
r (\mathbf{p}_1, \mathbf{p}_2) =
\!\!\int c(\mathbf{x}_1, \mathbf{x}_2)  {{f(\mathbf{x}_1,\mathbf{p}_1)}\over n_1(\mathbf{x}_1)} 
{{f(\mathbf{x}_2,\mathbf{p}_2)}\over{n_1(\mathbf{x}_2})}
d\Gamma_1d\Gamma_2.
\label{eq:MomCorr}
\end{equation}
We identify $c(\mathbf{x}_1, \mathbf{x}_2)$ at freeze out with (\ref{eq:param}), a form that describes the system at its formation. This identification omits the effects of diffusion described in ref.~\cite{gavin06}. This omission is reasonable only as long as we restrict our attention to the long range correlations with pairs separated by  $|\eta_1 - \eta_2| > 1$. 

STAR measures the characteristics of the untagged near-side ridge as functions of the centrality at 62 and 200 GeV for Au+Au \cite{Daugherity:2006hz}. While they focus on the region $-1 <\eta < 1$ where short and long range correlation phenomena are both present, it is instructive to see which aspects of the data can be explained by a purely long range model. To facilitate or comparison, we visualize the STAR analysis as consisting of the following steps. First, a correlated two particle distribution of ``sibling'' particles $\rho_{sib}$ is measured. This quantity is essentially our $\rho_2$ integrated over the magnitudes of each particle's $p_t$ as well as the average azimuthal angle $\Phi = (\phi_1 + \phi_2)/2$ and pseudorapidity   $\eta_a = (\eta_1 + \eta_2)/2$. The resulting density depends only on the relative quantities $\phi = \phi_1 - \phi_2$  and $\eta = \eta_1 - \eta_2$.  Second, an uncorrelated pair distribution $\rho_{ref}$ is obtained from mixed events and $\Delta \rho/\sqrt{\rho} = (\rho_{sib} - \rho_{ref})/\sqrt{\rho_{ref}}$ is constructed.  Next, a rapidity-independent function  $a + b \cos\phi + c \cos 2\phi$ is subtracted to remove backgrounds as well as elliptical flow and momentum conservation effects. Finally, the corrected $(\eta, \phi)$ distribution is  subjected to a multicomponent fit to extract the attributes of the near side peak. In practice, these steps are performed simultaneously.

To confront the STAR measurements, we calculate $\Delta \rho(\eta,\phi)$ by  integrating (\ref{eq:MomCorr}) over all momenta. Similarly, we compute $\rho_{ref}(\eta,\phi)$ by integrating $ \rho_1(\mathbf{p_1}) \rho_1(\mathbf{p_2})$.  The integrations are straightforward if we take the $\gamma$ factors as constants evaluated at $R_t = R_A$. We obtain
\begin{equation}
\Delta \rho/\sqrt{\rho_{ref}} ={\cal R} dN/dy\,\, F(\phi),
\label{eq:deltaRho}
\end{equation}
where $\int_{0}^{2\pi} F(\phi) d\phi =1$.
The factor 
\begin{eqnarray}
{\cal R}dN/dy= \kappa\alpha_s(Q_s)^{-1}
\label{eq:CGCscale}
\end{eqnarray}
follows from the Glasma relations (\ref{eq:Nscale}) and (\ref{eq:Rscale}), 
where $\kappa$ is an energy independent constant to be determined from data. The angular distribution $F(\phi)$ depends only on blast wave parameters $\gamma m/T$ and $v_s$.

We specify the centrality dependence of the correlation function using the velocity and temperature fit from single particle spectra at 200 GeV in ref.~\cite{Kiyomichi:2005va}. The computed $\Delta\rho/\sqrt{\rho}$ is shown as the dashed line in the top panel of fig.~1. We fix $\kappa$ in (\ref{eq:CGCscale}) and (\ref{eq:deltaRho}) to agree with the magnitude of the 200 GeV data. We define the height of the near side peak as the difference between $\Delta\rho/\sqrt{\rho}$ at $\phi = 0$ and $\pi$. The dashed line in the top panel is the blast wave result $F(\phi)$ without the CGC scaling. It follows the basic trend of the data rather well, given that the only parameters that vary with centrality  -- $v$ and $T$  -- are fit elsewhere \cite{Kiyomichi:2005va}. The uncertainty in $v$ and $T$ implies the shaded bands in fig.~1. 
To compute  $\Delta\rho/\sqrt{\rho}$ at 62 GeV, we follow \cite{Kiyomichi:2005va} and reduce the velocities by $5\%$ and the temperatures by $10\%$ by uniform scale factors. The dashed curve in the bottom panel of fig.~1 is well above the data, but agrees roughly in shape.  
\begin{figure}
\centerline{\includegraphics[width=3.2in]{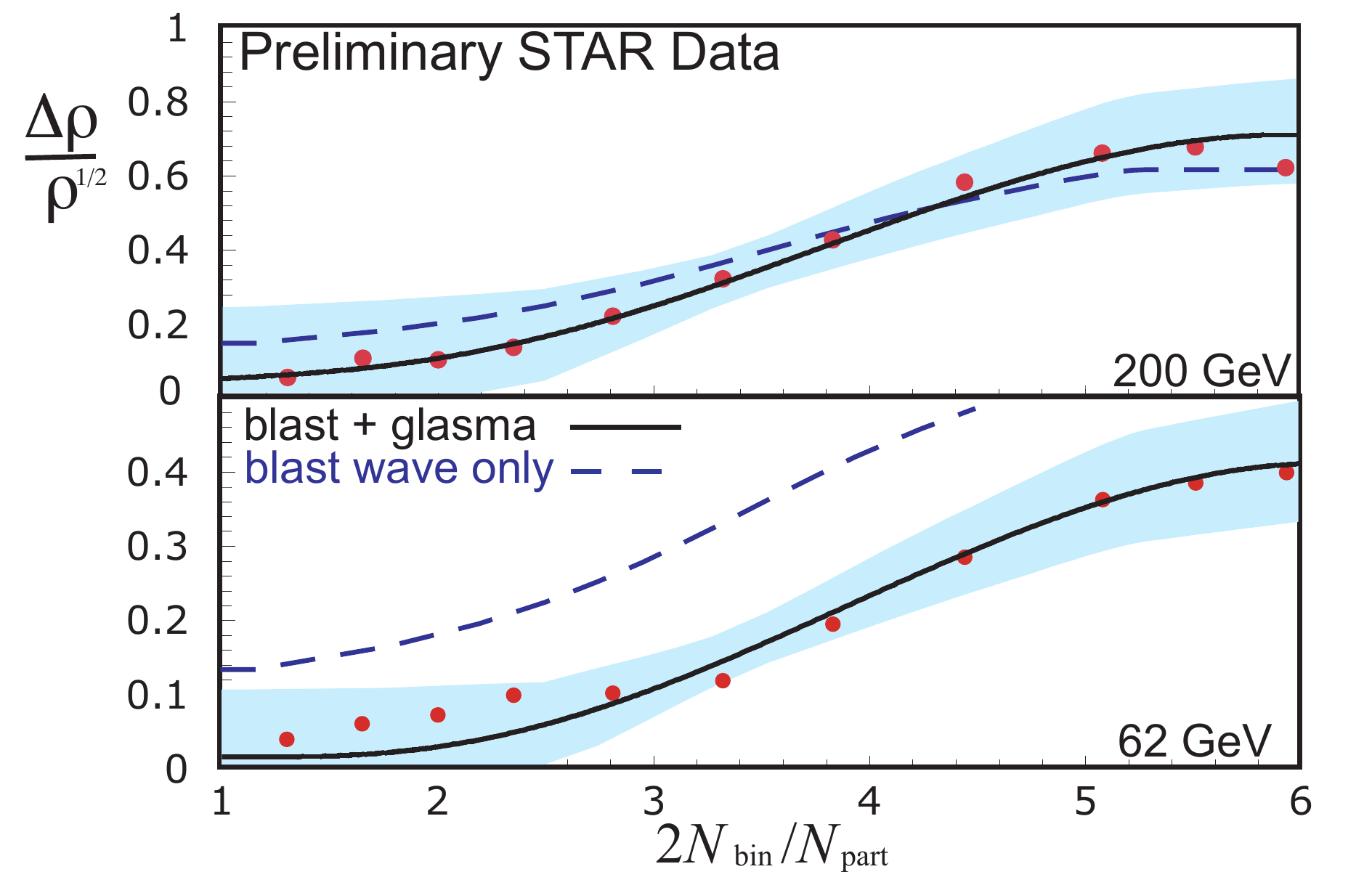}}
\caption[]{Height of the near side peak vs.\ centrality for 200 GeV (top) and 62 GeV (bottom). Preliminary STAR data is from \cite{Daugherity:2006hz}.  Bands indicate the uncertainty of the blast wave parameters $T$ and $v$.}
\label{fig:fig2}\end{figure}

The very existence of long range correlations implies strong correlations in the initial state, as predicted by Color Glass Condensate theory \cite{larry}. That said, it is interesting to see how (\ref{eq:CGCscale}) influences the systematics of the soft ridge. This modifies the centrality dependence of $\Delta \rho/\sqrt{\rho}$ by introducing a logarithmic dependence on $Q_s$, which in turn depends on $N_{part}$. It is important to note that although the dashed curve represents the blast wave without the $\alpha_s$ scaling, the correlation function in the blast wave integrals enforces restraints from the Glasma picture. The upper solid curve in fig.~1 combines this $\alpha_s(N_{part})^{-1}$ dependence from ref.~\cite{Kharzeev:2000ph} with the  blast wave behavior. Agreement with data is impressive given that $Q_s(N_{part})$ is obtained fit in \cite{Kharzeev:2000ph}.  

We deduce $\Delta\rho/\sqrt{\rho}$ at 62 GeV using the $Q_s$ dependence discussed earlier together with the relevant blast wave $v$ and $T$. With the proportionality constant in (\ref{eq:CGCscale}) fixed by the 200 GeV data, there are no further free parameters to adjust. The solid curve in the lower panel on fig.~1 is in good accord with the data for the expected drop of $Q_s^2$ by $\sim 1/2$ relative to the 200 GeV value \cite{Kharzeev:2000ph}. Observe that most of the change from 62 to 200 GeV owes to the  $\alpha_s^{-1}(Q_s)$ dependence, since the change in the blast wave parameters is small \cite{Iordanova:2007vw}. 
 
We compare our calculations to the measured azimuthal width of the near side peak in fig.~2. To simulate the experimental fit procedure, we obtain this width by fitting a gaussian plus a constant offset to the computed $\phi$ distribution in the near-side interval $-\pi/2 < \phi < \pi/2$. The uncertainty band in fig.~2 indicates the impact of changing the near-side interval by $\pm 20\%$. Once again, the agreement is surprisingly good given the simplicity of the model.  The calculated angular width does not change in this energy range, since the normalization does not affect $F(\phi)$.  Note that ref.~\cite{larry} includes flow by boosting a Glasma source. Their computed width exceeds data because they omit the $p_t$ dependence of the boost and do not simulate the experimental fit procedure. 
\begin{figure}
\centerline{\includegraphics[width=3.2in]{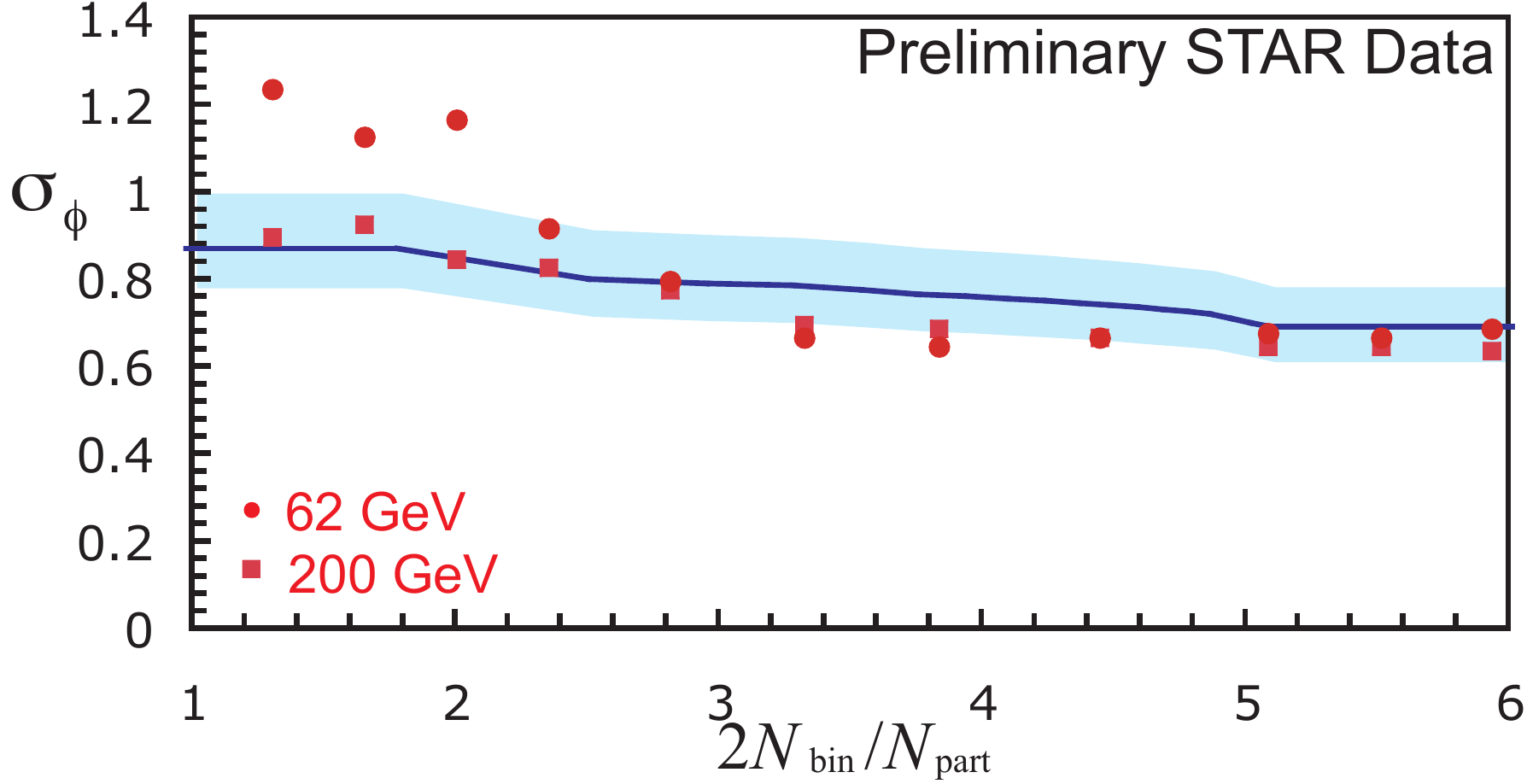}}
\caption[]{Azimuthal width of the near side peak vs. centrality. Preliminary STAR data is from \cite{Daugherity:2006hz}. The curve is obtained for by fitting a gaussian to the computed distribution in the range $-\pi/2<\phi < \pi/2$. The band shows the sensitivity of the result to a 20\% change in this range.}
\label{fig:fig1}\end{figure}

In summary, correlation measurements show a ridge that is narrow in $\phi$ and broad in $\eta$, perhaps extending several units in rapidity. Such long-range rapidity correlations can only be caused by super-horizon fluctuations at sub-fermi time scales. While flux tubes may also enhance forward-backward correlations \cite{Armesto:2006bv}. These measurements therefore provide an 
image of the particle production process at the sub-fermi scale, which can be corroborated e.g., by forward-backward correlation measurements \cite{Armesto:2006bv}.  For simplicity we have focused on the soft ridge, but similar considerations may also apply to jet-tagged measurements \cite{Shuryak:2007fu}. 
The height and azimuthal width measured near midrapidity in ref.~\cite{Daugherity:2006hz} are consistent with flux tubes of large longitudinal extent formed early in nuclear collisions. Correlations predicted by Color Glass Condensate theory combined with transverse flow provide a remarkably good description of the near side ridge. In particular, agreement with 62 GeV data follows mainly from the CGC-Glasma prediction (\ref{eq:Rscale}), the only free parameter being an overall constant fixed at 200 GeV.  This agreement is surprising, since our model only includes  long range correlations. The rapidity  dependence for  $|\eta| < 1 - 2$ requires a more detailed hydrodynamic description \cite{gavin06}. 

S.G. thanks the nuclear theory groups at Brookhaven and University of Minnesota for hospitality. We thank M. Baker, R. Bellwied, C. De Silva, A. Dumitru, F. Gelis,  J. Kapusta, L. Ray, T. Springer, P. Sorenson, P. Steinberg, R. Venugopalan, and S. Voloshin.  This work was supported in part by U.S. NSF PECASE/CAREER grant PHY-0348559 (S.G. and G.M.) and U.S. DOE Contract No.~DE-AC02-98CH10886 (L.M.).

\vfill\eject
\end{document}